**ЄЧКАЛО Ю. В.**, доцент кафедри фізики ДВНЗ «Криворізький національний університет»

# МЕТОДИ НАВЧАННЯ КОМП'ЮТЕРНОГО МОДЕЛЮВАННЯ ФІЗИЧНИХ ПРОЦЕСІВ І ЯВИЩ У ВИЩІЙ ШКОЛІ

*У статті виконано аналіз методів навчання комп'ютерного моделювання фізичних процесів і явищ у вищій школі. Основними методами навчання комп'ютерного моделювання у вищій школі є мультимедійна лекція, телекомунікаційний проект та комп'ютерно-орієнтований лабораторний практикум.*

**Ключові слова**: *метод навчання, комп'ютерне моделювання.*

**Постановка проблеми**. Удосконалення традиційних форм вищої освіти та пошук нових підходів, ідей та методів навчання, здатних покращити якість освіти та рівень підготовки випускників, зокрема з фізики (яка є фундаментальною основою технічних дисциплін), – актуальна проблема сучасної інженерної освіти. Для ефективної підготовки студентів інженерних спеціальностей потрібне формування системи фундаментальних фізичних знань в сукупності з уміннями застосовувати їх в конкретній виробничій діяльності, як на фундаментальному, так і на профільно-орієнтованому рівні. Відповідно до цього завданнями курсу фізики вищої школи є, зокрема, оволодіння студентами методологією природничо-наукового пізнання і науковим стилем мислення, узагальнене експериментальне вміння вести природничо-наукові дослідження методами фізичного пізнання. Застосування наукового методу у навчанні фізиці відіграє подвійну роль: з одного боку, він є об'єктом вивчення, з іншого – виступає у якості ефективного методу повідомлення знань.

**Аналіз останніх досліджень і публікацій**. Методи навчання – впорядковані способи взаємопов'язаної діяльності викладача та студента (їх взаємосприяння), спрямовані на досягнення цілей навчання. За методом навчання визначається, що і як саме студенти повинні робити з навчальним матеріалом, які властивості і зв'язки між об'єктами необхідно розкривати. Метод є центральною ланкою детермінації процесу навчання зовнішніми обставинами. Поряд з поняттям «метод навчання» у теорії й педагогічній практиці використовуються поняття «прийом навчання», «методичний прийом». Прийнято вважати, що метод як спосіб діяльності складається із прийомів або окремих дій, спрямованих на розв'язування педагогічних завдань [1; 2].

У методах навчання можна виділити змістову і формальну сторони. Змістова сторона включає такі компоненти:

1) зміст, різні моделі, аналогії, алгоритми, використання яких дає змогу засвоїти сутність виучуваних об'єктів;

2) розумові, передусім мислительні, дії, потрібні для засвоєння змісту об'єктів навчання і додаткового змісту (загальнологічні дії, а також дії, через які розкриваються принципи побудови навчального матеріалу тощо);

3) співвідношення між цілями навчання, з одного боку, та прямими і непрямими його продуктами з іншого.

Формальна сторона методів навчання характеризується співвідношенням активності викладача та студентів, характером поєднання колективних та індивідуальних форм навчальної роботи, співвідношенням зорових та слухових форм подання навчального матеріалу, кількістю і складністю завдань, які стоять перед студентами, мірою допомоги, що надається їм тощо. При цьому діяльність викладача, з одного боку, обумовлена метою навчання, закономірностями засвоєння й характером навчальної діяльності студентів, а з іншого боку – вона сама обумовлює діяльність студентів, реалізацію закономірностей засвоєння й розвитку [3].

Оскільки загальні методи навчання численні й мають багато характеристик, їх можна класифікувати за кількома напрямами:

1. За характером спільної діяльності викладача та студентів: репродуктивний метод, пояснювально-ілюстративний метод, метод проблемного подання навчального матеріалу, частково-пошуковий або евристичний метод, дослідницький метод тощо [4].

2. За основними компонентами діяльності викладача: методи організації й здійснення навчальної діяльності, методи стимулювання й мотивації навчання, методи контролю й самоконтролю [5].

Частково-дидактичні методи навчання можна класифікувати:

– за особливостями подання та характером сприймання матеріалу: словесні методи (розповідь, бесіда, лекція та ін.); наочні (показ, демонстрація та ін.); практичні (лабораторні роботи, твори та ін.) [6; 7];

– за ступенем взаємодії викладача та студентів: подання матеріалу, бесіда, самостійна робота;

– в залежності від конкретних дидактичних завдань: підготовка до сприймання, пояснення, закріплення матеріалу й т.д.;

– за принципом розчленовування або з'єднання знань: аналітичний, синтетичний, порівняльний, узагальнюючий, класифікаційний;

– за характером руху думки від незнання до знання: індуктивний, дедуктивний.

У процесі навчання фізики моделювання одночасно виступає методом наукового пізнання, є частиною змісту навчального матеріалу та ефективним засобом його вивчення.

**Мета статті** – аналіз методів навчання комп'ютерного моделювання фізичних процесів і явищ у вищій школі.

**Виклад основного матеріалу.** Крім загально-дидактичних та частково-дидактичних, виділимо спеціальні методи навчання комп'ютерного моделювання, до яких відноситься насамперед обчислювальний експеримент. Це пов'язано з такими обставинами:

1) обчислювальний експеримент є методологією моделювання як науки, тому його можна віднести до принципів (методології) наукових методів учіння;

2) цілі навчання фізики у вищій школі включають необхідність засвоєння як певної сукупності наукових фактів, так і методів отримання цих фактів, які використовуються в самій науці, а обчислювальний експеримент відображає метод пізнання, що застосовується у фізиці.

Відповідно до вищенаведеного, схема класифікації методів навчання комп'ютерного моделювання може мати вигляд, показаний на рис. 1 [3].

При виборі та поєднанні методів навчання необхідно керуватися наступними критеріями:

– відповідність цілям і завданням навчання, виховання й розвитку;

– відповідність змісту досліджуваного матеріалу (складність, новизна, характер, можливість наочного подання матеріалу);

– відповідність реальним навчальним можливостям студентів: рівню підготовленості (навченості, розвиненості, вихованості, ступінь володіння інформаційними й комунікаційними технологіями), особливостям групи;

– відповідність наявним технічним умовам та відведеному для навчання часу;

– відповідність індивідуальним особливостям і можливостям самих викладачів (риси характеру, рівень володіння тим чи іншим методом, стосунки з групою, попередній досвід, рівень психолого-педагогічної, методичної та інформаційно-технологічної підготовки).

Частина назв форм навчання виступають і в якості назв методів навчання: це, насамперед, лекція та метод проектів.

Основними методами навчання комп'ютерного моделювання у вищій школі є мультимедійна лекція, телекомунікаційний проект та комп'ютерно-орієнтований лабораторний практикум.

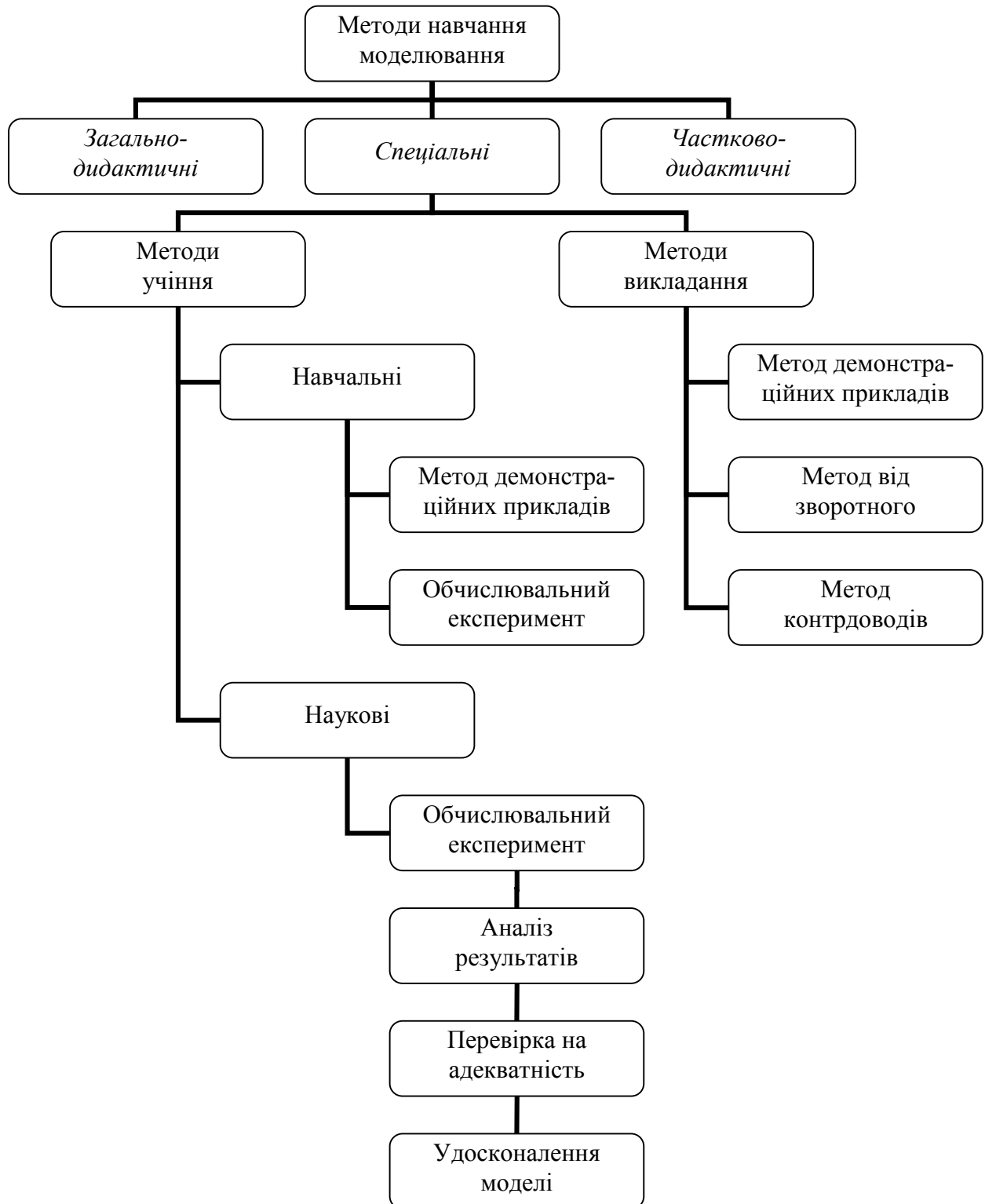

*Рис. 1. Класифікація методів навчання комп'ютерного моделювання*

Мультимедійна лекція – таке викладення навчального матеріалу, у якому лектор, передаючи комп'ютеру частину своїх функцій, посилює вплив на слухачів шляхом використання можливостей, що надаються йому мультимедійними технологіями [8, с. 43]. Мультимедійні лекції повинні:

– відповідати науковому рівню вимог, які пред'являються до лекцій;
– ефективно стимулювати навчально-пізнавальну діяльність студентів;
– оптимально візуалізувати навчальний матеріал;
– мати універсальність у виконанні, забезпечувати варіативність у поданні навчального матеріалу, відповідаючи практичним потребам викладача та студентів;
– раціонально поєднувати різні технології пред'явлення навчального матеріалу;
– розвивати інтелектуальний потенціал студентів;
– забезпечувати контроль знань.

У процес створення мультимедійних лекцій повинні входити:
– розробка педагогічного сценарію до мультимедійних лекцій;
– розробка комп'ютерного сценарію (підготовка тексту, ілюстрацій для мультимедійних лекцій, вибір технологій та інструментальних засобів);
– безпосереднє створення мультимедійних лекцій та їх застосування в навчально-виховному процесі [8].

До підготовки мультимедійних лекцій можуть бути активно залучені студенти. Така форма співпраці викладача та студентів становить основу діяльнісного методу навчання, коли студент отримує не тільки знання, але також конкретні уміння при виконанні суспільно-корисної роботи.

Метод проектів, незважаючи на свої давні витоки – один з основних сучасних інноваційних методів активного навчання. В навчанні фізики цей метод широко впроваджується в освітню практику, теоретичні основи впровадження методу проектів розроблені в працях Є. С. Полат. Проекти можуть бути індивідуальними і груповими, локальними та телекомунікаційними. Під навчальним телекомунікаційним проектом Є. С. Полат розуміє таку форму навчання, яка передбачає спільну навчально-розвивальну діяльність учасників, які можуть бути територіально віддаленими, для досягнення значущої для них мети (результату) узгодженими методами, що вимагають застосування засобів комп'ютерних телекомунікацій [9]. Характерними ознаками навчальних телекомунікаційних проектів є самостійна дослідницька діяльність їх учасників, пов'язана з розв'язанням певної проблеми, що має на меті отримання практичного результату та спирається на більшості або на кожному своєму етапі на використання засобів комп'ютерних телекомунікацій.

Враховуючи різні підходи до класифікації проектів у педагогічній літературі, пропонується розрізняти їх за цілим рядом параметрів: кількістю учасників проектної діяльності (індивідуальні, колективні – парні та групові); характером партнерських взаємодій між учасниками проектної діяльності (кооперативні, змагальні, конкурсні); мірою реалізації міжпредметних зв'язків (монопредметні, міжпредметні, позапредметні або надпредметні); характером координації проекту (безпосередній або жорсткий чи гнучкий, прихований); тривалістю (короткотривалі, довготривалі); метою та характером проектної діяльності (інформаційні, творчі, прикладні, дослідницькі тощо). Внутрішня структура проекту вимагає наявності традиційних компонентів: актуальність проблеми, предмет дослідження, мета проекту, гіпотези, задачі, використані методи, практична значущість результату. Однак оскільки навчальні проекти бувають різні, в залежності від характеру діяльності, яка лежить в основі того чи іншого проекту, то і методика виконання конкретного проекту буде дещо відрізнятися [10].

Дослідницькі проекти, до яких можна віднести створення комп'ютерних моделей,

мають чітко визначене дослідницьке завдання, повністю підпорядковані загальній логіці та мають структуру, наближену до структури наукового дослідження або таку, яка повністю співпадає з нею, а саме: аргументація актуальності теми, що прийнята для дослідження, виділення проблеми та мети дослідження, формулювання гіпотези дослідження, визначення методів дослідження, джерел інформації, обговорення, аналіз і оформлення отриманих результатів. Структура дослідницького проекту зі створення комп'ютерної моделі представлена у таблиці 1 [11].

**Таблиця 1**

*Етапи виконання дослідницького проекту зі створення комп'ютерної моделі*

| Етапи проекту | Зміст етапу |
|---|---|
| Визначення мети | Створення й дослідження комп'ютерної моделі фізичного явища чи процесу. |
| Очікуваний результат | Не завжди відомий на початку дослідження, висувається гіпотеза про результати, яка потім піддається експериментальній або теоретичній перевірці. |
| Розробка моделі | Розглядаючи у найбільш загальних рисах структуру процесу моделювання, визначають такі її складові:<br>– актуалізація знань про об'єкт-оригінал;<br>– вибір інформаційної моделі з числа існуючих або створення такої моделі;<br>– дослідження моделі;<br>– перенесення даних, що їх було одержано при дослідженні моделі, на оригінал;<br>– перевірка істинності даних, одержаних за допомогою моделі і включення їх до системи знань про оригінал. |
| Визначення форми продукту проектної діяльності | Результати роботи повинні бути представлені в чисельному вигляді, у вигляді графіків, діаграм. Якщо є можливість, процес представляється у динаміці. Після закінчення розрахунків і одержання результатів проводиться їхній аналіз, порівняння з відомими фактами з теорії, підтверджується вірогідність і проводиться змістовна інтерпретація, що надалі відбивається в письмовому звіті. Звіт містить короткі теоретичні відомості із досліджуваної теми, математичну постановку завдання, алгоритм розв'язання і його обґрунтування, результати роботи моделі, аналіз результатів і висновки, список використаної літератури. |
| Презентація проекту | Коли всі звіти складені, на заліковому занятті студенти виступають із короткими повідомленнями про виконану роботу, захищають свій проект. Це є ефективною формою звіту групи, що виконує проект, включаючи постановку завдання, побудову формальної моделі, вибір методів роботи з моделлю, реалізацію моделі на комп'ютері, роботу з готовою моделлю, інтерпретацію отриманих результатів. |
| Оцінювання | У підсумку студенти отримують дві оцінки: першу за глибину опрацювання проекту й успішність його захисту, другу – за модель, оптимальність її алгоритму, інтерфейс тощо. Також студенти отримують оцінки в ході опитувань з теорії. |
| Значущість проекту | Формування основ наукового світогляду, розвиток мислення та здібностей студентів, формування навичок дослідження, підготовка студентів до практичної діяльності. |

Метод навчальних проектів дозволяє внести в сучасну практику навчання два істотних доповнення – зміна у функції знань і способів організації процесу їхнього засвоєння. Процес засвоєння знань перестає носити характер рутинного заучування і організується в різноманітних формах пошукової, проектної, розумової діяльності як продуктивний творчий процес. Основою навчального проектування стає засвоєння як знань, так і способів їх добування, а також розвиток інтелектуальних здібностей студента [11].

Комп'ютерно-орієнтована лабораторна робота (фронтальна) є основною формою роботи в комп'ютерній аудиторії. Діяльність студентів може бути як синхронна, так і асинхронна. Нерідко відбувається швидке «розтікання» фронтальної діяльності навіть при спільному вихідному завданні. Роль викладача під час фронтальної лабораторної роботи – спостереження за роботою студентів (у тому числі через локальну мережу) та надання їм оперативної допомоги.

На таких заняттях студенти повинні розробити математичну модель явища або процесу, реалізувати її на комп'ютері, а потім виконати з такою моделлю ряд експериментів, як на звичайній лабораторній роботі. При цьому передбачається наявність знань теоретичного матеріалу, активне включення у творчу діяльність, що істотно збільшує результативність навчального процесу. За результатами експерименту з моделлю студенти мають зробити відповідні висновки.

Індивідуальний практикум – більш високорівнева форма роботи в порівнянні із фронтальними лабораторними роботами, що характеризується різнотипністю завдань як за рівнем складності, так і за рівнем самостійності; більшою опорою на підручники, довідковий матеріал, ресурси Інтернет тощо.

Студенти одержують індивідуальні завдання від викладача на одне, два або більше занять. Як правило, такі завдання видаються для відпрацьовування знань та вмінь, що відповідають певному розділу (модулю) курсу фізики.

Практичні заняття є перехідною формою від фронтальної до індивідуальної роботи. Практичне заняття – найбільш адекватна форма роботи для колективного осмислення того, що треба зробити або вже зроблено на комп'ютері, і чому такі результати отримані.

Важливим інтелектуальним умінням є здатність до розгорнутого прогнозу результатів, отриманих за допомогою комп'ютера на основі накопиченого досвіду роботи з ним. Для його формування доцільно застосовувати семінарські заняття. На семінарах з комп'ютерного моделювання можливе колективне обговорення:
– змісту моделі;
– тексту чи блок-схеми алгоритму реалізації моделі;
– таблиці виконання алгоритму (без застосування комп'ютера);
– плану та результатів обчислювального експерименту;
– нових версій моделі.

**Висновки**. Основними завданнями навчання комп'ютерного моделювання в курсі фізики є загальний розвиток і становлення світогляду майбутніх інженерів, оволодіння моделюванням як методом пізнання, вироблення практичних навичок комп'ютерного моделювання, розвиток і професіоналізація навичок роботи з комп'ютером, формування навичок проектної діяльності. Провідним методом навчання комп'ютерного моделювання в курсі фізики є метод проектів. Розглянуті методи навчання є оптимальними в контексті розвиваючого навчання.

**YECHKALO YU.,**
Associate professor, SIHE "Kryvyi Rih National University", Krivyi Rig

**METHODS OF LEARNING OF COMPUTER SIMULATION OF PHYSICAL PROCESSES AND PHENOMENA IN UNIVERSITY**

*Abstract. Introduction.* To effectively prepare engineering students requires of formation of a system of fundamental physical knowledge together with the ability to apply them in specific productive activities, both on fundamental and on the profiled-oriented level. Accordingly the tasks of physics of high school is mastering the methodology of science knowledge and scientific way of thinking and generalize of experimental natural ability to conduct scientific research by methods of physical knowledge.

*Purpose.* Analysis of the learning methods of computer simulation of physical processes and phenomena in university.

*Methods.* In the process of teaching physics simulations simultaneously acts by scientific knowledge, is part of the content of educational material and effective means of experiment. It is possible to obtain special teaching methods of computer simulation, which include primarily computer experiment. The main methods of computer modeling learning in university are a multimedia lecture, telecommunications project and computer-oriented laboratory practice.

*Results.* The main tasks of teaching computer modeling in physics course are the overall development and formation of outlook of future engineers and development practical skills in computer simulation. The leading method of learning computer simulation in physics course is a project method. Those methods of learning are optimal in the context of developmental education.

*Conclusion.* Computational experiment is a modeling methodology as a science, so it can be attributed to the principles of scientific methods of learning. Purposes of learning physics in university include the necessity of mastering a given set of scientific facts and methods of getting these facts. Computational experiment reflects the method of knowledge which applied in physics.

*Keywords*: methods of learning, computer simulation.